\documentclass[12pt]{article}
\usepackage{amsfonts,epsfig,latexsym,amscd,multicol,theorem,amssymb}

\newcommand{\R}{{\mathbb{R}}}
\newcommand{\Z}{{\mathbb{Z}}}

\newcommand{\C}{{\mathbb{C}}}

\newcommand{\CP}{{\mathbb{C}}{\mathbb{P}}}  

\newcommand{\beq}{\begin{equation}}
\newcommand{\eeq}{\end{equation}}
\newcommand{\bea}{\begin{eqnarray}}
\newcommand{\eea}{\end{eqnarray}}
\newcommand{\ra}{\rightarrow}

\newcommand{\cd}{\partial}

\newcommand{\M}{{\sf M}}

\newcommand{\half}{\frac{1}{2}}

\newcommand{\so}{{\mathfrak{so}}}
\newcommand{\e}{{\mathfrak{e}}}
\newcommand{\aut}{{\mathfrak{aut}}}
\newcommand{\A}{{\mathcal{A}}}
\newcommand{\B}{{\mathcal{B}}}
\newcommand{\X}{{\sf X}}

\newcommand{\U}{{\mathcal{U}}}

\newcommand{\ol}{\overline}

\newcommand{\xv}{{\bf x}}
\newcommand{\yv}{{\bf y}}

\newcommand{\re}{{\rm Re}\, }
\newcommand{\im}{{\rm Im}\, }
\newcommand{\spec}{{\rm spec}\, }

\theoremstyle{plain}
\newtheorem{thm}{Theorem}

\newtheorem{prop}[thm]{Proposition}

{\theorembodyfont{\rmfamily}

}
\newcommand{\news}{\setcounter{equation}{0}}

\renewcommand{\theequation}{\thesection.\arabic{equation}}

\begin{document}

\title{Slow Schr\"odinger dynamics of gauged vortices}
\author{
N.M. Rom\~ao\thanks{e-mail: {\tt nromao@mpim-bonn.mpg.de}}\\
Max-Planck-Institut f\"ur Mathematik\\
Vivatsgasse 7, D--53111 Bonn, Germany\\[.5cm]
J.M. Speight\thanks{
e-mail: {\tt j.m.speight@leeds.ac.uk}}\\
Department of Pure Mathematics, University of Leeds\\
Leeds LS2 9JT, England}

\date{}
\maketitle

\begin{abstract}
Multivortex dynamics in Manton's
Schr\"odinger--Chern--Simons variant of the 
Landau--Ginzburg model of thin superconductors is studied within
a moduli space approximation. It is shown that the reduced flow on $\M_N$,
the $N$ vortex moduli space, is hamiltonian with respect to $\omega_{L^2}$,
the $L^2$
K\"ahler form on $\M_N$. A purely hamiltonian discussion of the conserved
momenta associated with the 
euclidean symmetry of the model is given, and
it is shown that the euclidean
action on $(\M_N,\omega_{L^2})$ is not hamiltonian. It is argued that
the $N=3$ flow is integrable in the sense of Liouville. 
Asymptotic formulae for $\omega_{L^2}$ and the reduced Hamiltonian for
large intervortex separation are conjectured. Using these,
a qualitative analysis of internal 3-vortex dynamics is given and a spectral
stability analysis of certain rotating vortex polygons is performed. 
Comparison is made with the dynamics of classical fluid point vortices and
geostrophic vortices. 
\end{abstract}

\maketitle

\section{Introduction}
\label{sec:intro}
\news

The Landau--Ginzburg theory of an idealized planar superconductor consists
of a complex scalar field $\phi$ representing the electron pair condensate, and
a $U(1)$ gauge potential $A_i$ ($i=1,2$), interacting via the potential
energy functional
\beq
V=\int\left(\half 
B^2+\half\sum_i D_i\phi\ol{D_i\phi}+\frac{\mu^2}{8}(1-|\phi|^2)^2
\right)d^2\xv.
\eeq
Here $D_i\phi=\cd_i\phi-iA_i\phi$ is the gauge covariant derivative, 
$B=\cd_1A_2-\cd_2A_1$ is the magnetic field and $\mu$ is a  
coupling constant. The model admits topologically stable, spatially localized
solutions (minimals of $V$) called vortices, the planar analogues of magnetic
flux tubes in 3-dimensional superconductors. 
These have finite $V$, and so have a 
well-defined winding number at infinity,
$N\in\Z$, which is interpreted physically as the net vortex number.
The case of critical coupling,
$\mu=1$, is special because here, given any unordered choice of $N$ complex
numbers (not necessarily distinct) $\{z_1,\ldots, z_N\}$, there exists a
minimal of $V$ with $N$ vortices located at $z=x_1+ix_2=
z_r$, $r=1,\ldots,N$,
unique up to gauge equivalence \cite{jaftau}. More precisely the Higgs field
$\phi$ of the solution has zeroes precisely at the prescribed points $z_r$.
Physically, one says that static critically coupled vortices exert no net
forces
on one another. It follows that the moduli space of static $N$-vortex
solutions is $\M_N\equiv\C^N$. Note that the vortex positions $\{z_r\}$
are not good global coordinates on $\M_N$ because we must identify solutions
which differ by permutation of $\{z_r\}$. Good global coordinates are
provided by the coefficients $w_r$ of the unique monic polynomial in $z$ 
whose roots are $\{z_r\}$ (counted with multiplicity), that is
\beq
\label{wdef}
z^N+w_1z^{N-1}+\cdots+w_N:=(z-z_1)\cdots(z-z_N).
\eeq

To introduce dynamics to the theory, one must also define a kinetic energy
functional, $T$, and many choices are possible. Manton has advocated the
use of a Schr\"odinger--Chern--Simons functional linear in first time 
derivatives, explicitly,
\beq 
\label{lagran}
T=\gamma\int\left(
\frac{i}{2}(\ol{\phi}D_0\phi-\phi\ol{D_0\phi})+BA_0+E_2A_1-E_1A_2
-A_0
\right)d^2\xv
\eeq
where $A_0$ is the temporal gauge field, $E_i=\cd_iA_0-\cd_0A_i$ is the 
electric field and $\gamma$ is another coupling constant \cite{man}. The
resulting dynamics (governed by Lagrangian $L=T-V$) is first order in time
and non-dissipative, and is hoped to give a description of vortex dynamics
in thin superconductors at very low temperatures. 
The Lagrangian (\ref{lagran}) has also been related to effective 
theories of the fractional quantum Hall effect in \cite{romthe} and 
\cite{tonhall}.
Two different interpretations for the statistics of the solitons in this
model were given in \cite{hanetal} and \cite{rom}.

In order to make progress on the problem of $N$-vortex dynamics in this model,
one could resort to numerical solution of the field equations. Manton 
\cite{man} has, however, pursued a different strategy, namely that of
adiabatic approximation. One assumes that at each time the field configuration
lies in $\M_N$, the moduli space of static solutions of the $\mu=1$ system, but
that the field's position in $\M_N$ may vary slowly with time. This slow
variation is assumed to be governed by the reduced Lagrangian, that is,
$T-V$ restricted to $\M_N$.  
The result is a finite dimensional dynamical
system, in fact, a hamiltonian flow on $\M_N$, as we shall see, which
hopefully captures the important features of the full, infinite dimensional
field dynamics. Of course if $\mu=1$ exactly, the restricted
potential is constant, and there is no dynamics --- the vortices remain
static. On the other hand,
if $\mu$ differs greatly from $1$ then approximating 
multivortex profiles by critical $N$-vortices is ill-justified. 
So the approximation is valid only in the near critical regime. 
This strategy has proved to be extremely successful in the
study of vortex dynamics in the relativistic version of the Landau--Ginzburg
model \cite{sam,sha}, and has been justified by rigorous error estimates
\cite{stu1}. 

The present paper gives an analysis of vortex dynamics within the adiabatic
approximation, following on from Manton's original paper \cite{man}.
Some of the results are anticipated by the study of the equivalent
model on the two-sphere \cite{rom}. In section \ref{redsys}, for example, we
prove that the reduced flow on $\M_N$ is hamiltonian with respect to the
$L^2$ K\"ahler form, a result which 
can be understood from the results on $S^2$, at least formally, 
once the sphere's radius 
is taken to infinity. It is worthwhile giving independent proofs of
the relevant results for two reasons, however. First, the planar model has far
more direct physical significance, both as a model of extremely thin
superconductors, and of the 3-dimensional case 
where translation symmetry is imposed.
The spherical model is useful to study the quantum mechanics of the system,
but its direct advantage is mathematical convenience 
($\M_N \equiv \CP^N$ is then compact, hence the methods of geometric 
quantization are amenable to more explicit treatment) 
rather than physical applicability. 
Second, the formulation of the model on $S^2$
entails considerable technical subtleties absent in the planar model. When 
dealing with the planar model, therefore, many of the arguments simplify
considerably. 

Neither \cite{man} nor \cite{rom} contains a quantitative discussion of 
classical vortex dynamics, and for good reason --- neither the K\"ahler form
nor the Hamiltonian on $\M_N$ are explicitly known. In the second half of this 
paper we will restrict attention to the regime of well-separated vortices,
where explicit formulae can be inferred, and hence more progress is possible. 
The results turn out to be similar to the dynamics of so-called geostrophic
vortices \cite{geo}, of interest in meteorology, and, to a lesser extent,
classical fluid point vortices \cite{flu}. We give a complete description 
of the internal $3$-vortex dynamics and analyze the 
spectral stability properties of rigidly rotating regular polygons.

Lange and Schroers have studied the slow Schr\"odinger dynamics of {\em 
ungauged} Landau--Ginzburg vortices within an adiabatic approximation
\cite{lansch}. For them, there is no concept of critical coupling
related to a self-duality structure, 
so no moduli space of static $N$-vortex solutions is available. Instead, they
define $\M_2$ to be the unstable manifold of the coincident double vortex,
which they construct via numerical gradient flow. Their emphasis is very
different from the present work, therefore. They use intensive numerical
techniques to study two-vortex dynamics thoroughly, concentrating particularly
on the case of overlapping vortices.
Indeed, one of their main motivations was to provide the first 
(necessarily numerical)
implementation of the unstable manifold method for solitons in spatial 
dimension exceeding one. In contrast, we focus on the opposite
physical regime (large separation), for $N\geq 3$, where heavy
numerical analysis
is not required. 

\section{The reduced system}
\news
\label{redsys}

Let $q^\alpha$, $\alpha=1,\ldots,2N$, 
be some choice of local coordinates on
$\M_N$ (for example the real and imaginary parts of the vortex positions $z_r
\in\C$). Then, since $T$ depends only linearly on time derivatives, the
Lagrangian restricted to $\M_N$ must take the form
\beq
L|_{\M_N}= \A_\alpha(q)\dot{q}^\alpha-V(q)
\eeq
where $\A_\alpha$ are some functions of the coordinates $\{q^\beta\}$ which
may naturally be interpreted as the components of a $U(1)$ connexion form 
$\A=\A_\alpha dq^\alpha$ on
$\M_N$, since changing $\A$ by an exact form $\A\mapsto \A+d\Phi$ changes
$L|_{\M_N}$ by an irrelevant total time derivative \cite{man}. The 
Euler--Lagrange equations resulting from $L|_{\M_N}$ are
\beq
\label{*}
\B_{\alpha\beta}\dot{q}^\alpha=-\frac{\cd V}{\cd q^\beta},
\eeq
where $\B=d\A$ is the curvature form of $\A$. 
Clearly, $\B$ is closed. Further,
assuming that (\ref{*}) is a well-defined flow on $\M_N$, $\B$ must also
be a {\em nondegenerate} bilinear form. So we may reinterpret $\B$ as a 
symplectic form on $\M_N$. 
Given any function $F\in C^\infty(\M_N)$, let $\X_F$ denote its symplectic
gradient, that is the unique vector field satisfying
\beq
\iota_{\X_F}\B=-dF
\eeq
where $\iota$ denotes interior multiplication, $\iota_{X}\B(Y):=\B(X,Y)$.
System (\ref{*}) is then hamiltonian flow along
$\X_V$.

No explicit formula for $\B$ or $\A$ exists. However, Manton has shown how
$\A$ can be related to the analytic properties of the Higgs field $\phi$ near
its zeroes. Let us assume that the vortex positions $z_r$ are all distinct,
and use these as local complex coordinates on $\M_N$. Then it is known that
in a neighbourhood of $z_r$
\beq
\label{abdef}
\log|\phi(z)|^2=\log|z-z_r|^2+a_r+\half\ol{b}_r(z-z_r)+\half b_r(\ol{z}
-\ol{z}_r)+\cdots
\eeq
where $a_r$ and $b_r$ are some unknown functions of the vortex positions
$z_s$, $a_r$ real, $b_r$ complex.
We have used here the conventions of \cite{manspe}. 
Samols \cite{sam}, adapting earlier work of 
Strachan \cite{str}, showed that the $L^2$ metric on $\M_N$ could be written
entirely in terms of the unknown coefficients $b_r$, namely
\beq
\label{gL2}
g_{L^2}=\pi\sum_{r,s=1}^N\left(\delta_{rs}+2\frac{\cd b_s}{\cd z_r}\right)
dz_r d\ol{z}_s.
\eeq
This is a hermitian metric on $T\M_N$ which can be shown to be 
K\"ahler \cite{sam}.
Using similar techniques, Manton has obtained \cite{man}
a similar formula for the 
connexion form $\A$; in our notation,
\beq
\label{A1}
\A= i\pi\gamma\sum_{r=1}^N\left[(\ol{b}_r+\half\ol{z}_r)dz_r-(b_r+\half z_r)
d\ol{z}_r\right].
\eeq
We will now establish the following proposition:

\begin{prop} The connexion $\A$ has curvature form
$
\B=-2\gamma\omega_{L^2}
$
where $\omega_{L^2}$ is the K\"ahler form corresponding to $g_{L^2}$.
\end{prop}

\noindent
{\it Proof:} We will prove the formula on $\M_N\backslash\Delta_N$, where
$\Delta_N$ is the measure zero set on which vortices coincide, and appeal to
smoothness.
Let us define the $(0,1)$-form
\beq
\label{bform}
b:=\sum_r b_rd\ol{z}_r.
\eeq
Hermiticity of $g_{L^2}$ and (\ref{gL2}) imply
\beq
\label{id1}
\frac{\cd b_s}{\cd z_r}\equiv\frac{\cd\ol{b}_r}{\cd\ol{z}_s}\quad
\Rightarrow\quad \bar{\cd}\, \ol{b}=-\cd b.
\eeq
Note that the connexion $\A$ and $\omega_{L^2}$ may be 
written, using (\ref{bform}) and (\ref{id1})
\bea
\A&=& i\pi\gamma\left[\half\sum_r(\ol{z}_rdz_r-z_rd\ol{z}_r)+\ol{b}-b\right],
\nonumber \\
\omega_{L^2}&=&\frac{i \pi}{2}\left(\sum_r dz_r\wedge d\ol{z}_r +
\cd b - \bar{\cd}\, \ol{b}\right).
\eea
It follows that
\beq
2\omega_{L^2}+\frac{1}{\gamma}d\A=i\pi
(\cd b - \bar{\cd}\,\ol{b}+d\ol{b}-db)=i\pi(\cd\ol{b}-\bar{\cd}\,b).
\eeq
Now $\cd\bar{\cd}b=-\bar{\cd}\cd b=\bar{\cd}^2\ol{b}=0$ by (\ref{id1}), so
$\bar{\cd}b$ is an antiholomorphic $(0,2)$-form. Hence its component functions
are antiholomorphic functions on $\M_N\backslash\Delta_N$. But it is known
\cite{sam} that the coefficients $b_r$, and their derivatives,
decay exponentially fast at large vortex separation.
Hence these antiholomorphic functions must in fact vanish.
So $\bar{\cd}b=0=\cd\ol{b}$, and the proposition follows.\hfill$\Box$

For future reference, we note that $\bar{\cd}b=0$ implies that
\beq
\label{id2}
\frac{\cd b_r}{\cd \ol{z}_s}=\frac{\cd b_s}{\cd \ol{z}_r}
\eeq
for all $r,s$.

So the adiabatic approximation to $N$-vortex dynamics  is
hamiltonian flow on $(\M_N,\omega_{L^2})$ with Hamiltonian
\beq
H=-\frac{1}{2\gamma}V|_{\M_N}.
\eeq
Henceforth $\X_F$ will denote the symplectic gradient of $F$ with respect to
$\omega_{L^2}$, rather than $\B$.

\section{Conservation laws}
\news
\label{conlaw}

We may now give a hamiltonian discussion of the conservation laws discovered
by Manton and Nasir \cite{mannas}. The natural action of the euclidean group
$E(2)\cong U(1)\ltimes\C$ on the physical plane $\C$ induces a 
$E(2)$-action on $\M_N$ by $(e^{i\theta},c):\{z_r\}\mapsto\{e^{i\theta}z_r+c\}
$ (strictly speaking, this only defines the action on $\M_N\backslash\Delta_N$,
but one can use (\ref{wdef}) to deduce a well defined action on the global 
coordinates $w_r$). 
This action is manifestly holomorphic. It is also isometric, since it leaves 
$L^2$-norms invariant. Hence the $E(2)$-action 
is symplectic. We would like to construct a moment map $\mu:\M_N\ra
\e(2)^*$ for this action and identify its components with respect to a natural
basis for $\e(2)^*$ as the conserved momenta of the system, just as 
it has been done for the symplectic $SO(3)$-action in the $S^2$ 
model~\cite{rom}. Unfortunately, no such moment map exists.

Certainly, given any $X\in\e(2)\cong 
T_{(1,0)}E(2)$, the induced vector field
$X^\sharp\in\Gamma(T\M_N)$ is hamiltonian, because all symplectic vector
fields on $\M_N$ are hamiltonian 
($Y$ symplectic implies $\iota_Y\omega_{L^2}$ is closed;  
it is also exact since $H^1(\M_N)=0$, 
whence $Y$ is the symplectic gradient of some smooth function). 
Using the terminology of \cite{mcdsal}, the $E(2)$-action on $\M_N$ is 
{\em weakly hamiltonian}.
So functions $P_j:\M_N\ra\R$ $,j=0,1,2$ 
satisfying $-dP_j=\iota_{X_j}\omega_{L^2}$ must exist, where
\bea
X_0&=&\frac{\cd^\sharp}{\cd\theta}=i\sum_r\left(z_r\frac{\cd\, }{\cd z_r}-
\ol{z}_r\frac{\cd\, }{\cd\ol{z}_r}\right), 
\label{Xdef0} \\
X_1&=&\frac{\cd^\sharp}{\cd c_1}=\sum_r\left(\frac{\cd }{\cd z_r}+
\frac{\cd\, }{\cd\ol{z}_r}\right),\\
X_2&=&\frac{\cd^\sharp}{\cd c_2}=i\sum_r\left(\frac{\cd }{\cd z_r}-
\frac{\cd\, }{\cd\ol{z}_r}\right)
\label{Xdef2}
\eea
generate rotations and translations in 
$\M_N$. In fact, these functions are unique up to additive constants.
The problem is that the $P_j$ cannot be assembled into the components of
an {\em equivariant} map $\M_N\ra\e(2)^*$. In other words, we have:

\begin{prop} \label{actnotham}
The symplectic $E(2)$-action on $\M_N$ is not hamiltonian.
\end{prop}

\noindent
{\it Proof:}
Assume, to the contrary, that a moment map $\mu$ exists. Each $\M_N$ contains
a symplectic submanifold $(\C,\frac{i\pi N}{2}dZ\wedge d\ol{Z})$, namely the
$E(2)$-orbit of any configuration of $N$ coincident vortices (here $Z$ denotes
the common vortex position). The restriction $\hat{\mu}$ of $\mu$ to this
orbit defines a moment map for the standard action of $E(2)$ on $\C$. Since
$\hat{\mu}$ is equivariant, and the $SO(2)$ action fixes $0$,
\beq
\hat{\mu}(0)=\hat{\mu}((e^{i\theta},0)\cdot 0)=Ad^*_{(e^{i\theta},0)}
\hat{\mu}(0)
\eeq
where $Ad^*$ denotes the coadjoint action of $E(2)$ on $\e(2)^*$. Hence
$\hat{\mu}(0)=\alpha d\theta$ for some real constant $\alpha$.
But this completely determines $\hat{\mu}$ by transitivity of the 
translation
action:
\beq
\hat{\mu}(Z)=\hat{\mu}((1,Z)\cdot 0)=Ad_{(1,Z)}^*\alpha d\theta=\alpha d\theta.
\eeq
Thus, any equivariant map $\C\ra\e(2)^*$ is constant, and cannot generate
a nontrivial group action.\hfill$\Box$

Notwithstanding the lack of a moment map, generating functions 
$P_j$ for the $X_j$ do exist, and one 
expects, given Manton and Nasir's Lagrangian analysis \cite{mannas},
they may be
written locally (on $\M_N\backslash\Delta_N$) as
\bea
P_0&=&\frac{\pi}{2}\sum_r(|z_r|^2+b_r\ol{z}_r+\ol{b}_r z_r)\\
P_1&=&i\frac{\pi}{2}\sum_r(z_r-\ol{z}_r)\\
P_2&=&\frac{\pi}{2}\sum_r(z_r+\ol{z}_r).
\eea
Note that $P_1,P_2$ are easily globalized since $P_1+iP_2=
-i\pi w_1$ in the
global coordinates. A smooth globalization of $P_0$ must exist, but no
obvious formula suggests itself.

It is a routine exercise to 
verify that $dP_j(Y)=\omega_{L^2}(Y,X_j)$ for all $Y\in\Gamma(T\M_N)$
as required, if one makes use of the identities
\bea
\sum_r b_r&\equiv& 0 \label{samid}\\
\sum_r (b_r\ol{z}_r-\ol{b}_rz_r)&\equiv& 0\label{romid}.
\eea
Identity (\ref{samid}) is proved in \cite{sam}, while  (\ref{romid}) is
proved for the spherical model in \cite{rom}. For the sake of completeness,
we will derive it directly for the planar model.

Let $\phi\in\M_N$ be the Higgs field of the static solution with zeroes at
$\{z_r\}$, and let $\phi^\theta\in\M_N$ be the field obtained from this by
rotating all the vortex positions $z_r\mapsto e^{i\theta}z_r$. Then clearly
$\phi^\theta(e^{i\theta}z)\equiv\phi(z)$, and hence (\ref{abdef}) implies that
\beq
b_r(\{e^{i\theta} z_r\})=e^{i\theta}b_r(\{z_r\})
\eeq
for all $r$. It follows that
\beq
\label{deriv0}
X_0[b_r]=\left.\frac{d\, }{d\theta}\right|_{\theta=0}b_r=ib_r,
\quad\mbox{and}\quad 
X_0[\ol{b}_r]= -i\ol{b}_r,
\eeq
since $X_0$ is real. 
But using our local expression for $X_0$, (\ref{Xdef0}), we see that
\beq
\label{deriv}
X_0[b_s]=i\sum_r\left(z_r\frac{\cd b_s}{\cd z_r}-\ol{z}_r\frac{\cd b_s}{\cd
\ol{z}_r}\right)=
i\frac{\cd\, }{\cd\ol{z}_s}\sum_r(z_r\ol{b}_r-\ol{z}_r b_r)+ib_s
\eeq
where we have used (\ref{id1}) and (\ref{id2}). Comparing (\ref{deriv}) and
its complex conjugate with (\ref{deriv0}) one sees that 
$\sum_r(b_r\ol{z}_r-\ol{b}_rz_r)$ is constant. That the constant vanishes
follows from consideration of the limit where all $N$ vortices coincide. In
fact, constancy of the sum, rather than vanishing, is sufficient to prove
that $\X_{P_0}=X_0$. 

Our Hamiltonian 
$H=-\frac{1}{2\gamma}V|_{\M_N}$ is manifestly invariant under the
$E(2)$-action, so
\beq
\{P_j,H\}=\omega_{L^2}(X_j,\X_H)=dH(X_j)=X_j[H]=0
\eeq
and the momenta $P_j$ are conserved by the flow of $H$.

The Lie subalgebra generated by the $P_j$ in the Poisson algebra 
$C^{\infty}(\M_N)$ is determined by the relations
\bea
\{ P_j, P_0 \} & = &\epsilon_{jk}P_k, \qquad j=1,2\\
\{ P_1, P_2 \} & = & N\pi, \label{P1P2}
\eea
where $\epsilon_{jk}$ denotes the antisymmetric tensor with
$\epsilon_{12}=+1$. So we can see that it is a central extension
$\e(2)\oplus\R$ of the euclidean algebra. The nonvanishing of the Poisson
bracket (\ref{P1P2}) is consistent with a field theory calculation of
Hassa\"\i ne {\it et al.}~\cite{hashoryer}. In our context, given that 
$[X_1,X_2]=0$, it provides another proof of Proposition~\ref{actnotham}.
In fact, it implies that the $E(2)$-action on $\M_N$ corresponds to a nonzero
2-cocycle in the Lie algebra cohomology group $H^2(\e(2);\R)\cong\R$; this
group parametrizes obstructions of weakly hamiltonian $E(2)$-actions to be
hamiltonian \cite{mcdsal}.

It is worthwhile to look at  the content of Proposition~\ref{actnotham}
in the context of the results on the spherical model~\cite{rom}. In the
latter, spatial isometries are described by the group $SO(3)$, which is
simple. Therefore, $H^{2}(\so(3);\R)=0$ and the $SO(3)$-action
on $\M_N \equiv \CP^N$ is necessarily hamiltonian. Dualizing the corresponding
moment map, we obtain an isomorphism between $\so(3)$ and the 
Lie algebra of conserved angular momenta.
The large radius limit 
$R\ra \infty$ that relates the spherical model to the planar model determines 
a {\em contraction} of Lie algebras 
\beq
\so(3) \ra \e(2)
\eeq
with parameter $\frac{1}{R}$.
This is the well-known contraction of $\so(3)$ to the graded 
Lie algebra associated to the filtration \cite{guiste}
\beq
0 \subset \so(2) \subset \so(3),
\eeq
where the choice of the $\so(2)$ subalgebra corresponds to fixing an axis 
through the centre of the sphere.
We can realize this contraction as a deformation of Lie algebras of vector 
fields, using the isometry actions on each $\M_N$. At the level of the Poisson
algebras, however, no homomorphism of Lie algebras from $\e(2)$ to 
$C^{\infty}(\M_N)$ is available by Proposition~\ref{actnotham}, so it is not
surprising that the closure of the generators $P_j$ does not yield $\e(2)$ in
the planar model. A completely analogous discussion can be given for the 
zero-curvature limit relating the hyperbolic plane to the euclidean plane.

To end this section, we would like to emphasize that the fundamental reason 
underlying the nonhamiltonian action of the isometry group is the geometry 
of the domain of the model, rather than the dynamical nature of the model 
itself. Indeed, the proof of Proposition~\ref{actnotham} relies crucially 
on the fact that the standard 
action of $E(2)$ on the plane is not hamiltonian, so one would expect this
to be a common feature of symplectic moduli spaces of planar solitons. By 
contrast, the equivalent symmetry groups for the sphere and hyperbolic plane, 
$SO(3)$ and $SO(2,1)$ respectively, do have hamiltonian actions on the 
corresponding vortex moduli spaces. The lack of a moment map 
means that the standard Marsden-Weinstein technique for reducing the dynamics to
low-dimensional symplectic quotients is not available to us.
The Poisson noncommutativity of the components of the linear momentum (\ref{P1P2})
is
the ``classical anomaly'' which provides the obstruction. It 
is an expression of the fact that two of the real 
coordinates on phase space (essentially, the real and imaginary parts 
of $w_{1}$) are conjugate for the relevant symplectic structure, and should 
not be interpreted as a breaking of translational invariance.

\section{Well-separated vortices}
\label{welsepvor}
\news

Although the moduli space approximation has simplified the dynamical
$N$-vortex problem in principle, we are still faced with the problem that
neither the K\"ahler form $\omega_{L^2}$ nor the Hamiltonian 
$H=-\frac{1}{2\gamma} V$
are explicitly known. There is one physical regime in which explicit progress
is possible, namely  when all vortices are well separated from one another. In
this situation, an asymptotic formula for the coefficients $b_r$, and hence
$\omega_{L^2}$ is available \cite{manspe}. By translation invariance, the 
formulae involve only $z_{rs}:=z_r-z_s$. They are
\bea
\label{basymp}
& \displaystyle
b_r(z_1,\ldots,z_N)=\frac{q^2}{2\pi^2}\sum_{s\neq r}K_1(|z_{rs}|)
\frac{z_{rs}}{|z_{rs}|} & \\
\label{omasymp}
& \displaystyle
\omega_{L^2}=i\frac{\pi}{2}\left[\sum_r dz_r\wedge d\ol{z}_r -
\frac{q^2}{4\pi^2}
\sum_r \sum_{s\neq r}K_0(|z_{rs}|)dz_{rs}\wedge d\ol{z}_{rs}
\right], & 
\eea
where $K_n$ denotes the modified Bessel's function of the second kind, and
$q$ is an unknown real constant which may be interpreted as the scalar
monopole charge or magnetic dipole moment of a single vortex at critical 
coupling. Based on a string theoretic duality argument, Tong has
conjectured that $q=-2\pi 8^\frac{1}{4}$, and this value is consistent with
numerics \cite{ton,spe}. The formulae are believed to be correct up to linear
order in exponentially small quantities (note $K_n(\rho)$ is exponentially 
small for large $\rho$). They have not been proved rigorously, but they have
been derived via two independent physical arguments, one using matched
asymptotic expansions and the other appealing to a point source formalism,
and there is a good fit to known numerics.
Similarly derived formulae for monopole moduli spaces have subsequently
been proved rigorously, so 
one has good grounds for confidence.

It is not so straightforward to obtain an asymptotic formula for $H$. It is
known \cite{man} that
\beq
\label{v1}
V|_{\M_N}=N\pi+\frac{\mu^2-1}{8}\int(1-|\phi|^2)^2d^2\xv,
\eeq
but no obvious method of estimating the integral for 
well-separated vortices
suggests itself. We shall instead proceed by an indirect physical argument.
For $\mu$ close to $1$, one expects that $V|_{\M_N}$ is close to $V_N^\mu$,
the Landau--Ginzburg energy of $N$ 
well-separated coupling $\mu$ vortices held
at positions 
$\{z_r\}$, and for this an asymptotic formula {\em has} been
found \cite{spe}:
\beq
\label{4.4}
V_N^\mu=\frac{1}{2}
\sum_{r} \sum_{s\neq r}\tilde{\U}(|z_{rs}|)
\eeq
where $\tilde{\U}$ is the asymptotic 2-vortex interaction potential,
\beq
\label{Utdef}
\tilde{\U}(\rho):=\frac{1}{2\pi}[m(\mu)^2K_0(\rho)-q(\mu)^2K_0(\mu \rho)],
\eeq
and we have discarded an irrelevant constant ($N$ times the vortex mass).
The factor of $\half$ is included in (\ref{4.4}) since the sum is over all
{\em ordered} distinct vortex pairs.
Note that the scalar monopole charge $q(\mu)$ and magnetic dipole moment
$m(\mu)$ vary with the coupling, and are not equal except when $\mu=1$. Note
also that, at $\mu=1$, $\tilde{\U}\equiv 0$, as one expects. Now let $\mu=1+
\delta\mu$, $\delta\mu$ small, and assume that,
at sufficiently large vortex separation, $V|_{\M_N}=V_N^\mu$ to leading
order in $\delta\mu$. Again discarding additive constants, one is led to
conjecture that
\bea
V|_{\M_N}&=&\half
\sum_{r}\sum_{s\neq r}\U(|z_{rs}|)\nonumber \\
\label{Udef}
\U(\rho)&:=&\frac{q^2}
{4\pi}(\mu^2-1)\left[\rho K_1(\rho)-\nu K_0(\rho)
\right],
\eea
where $\nu=2(q'(1)-m'(1))/q\approx 2.7060$, using the tabulated data for
$m(\mu)$ and $q(\mu)$ in \cite{spe}. 
Note that equation (\ref{v1}) is an exact equality
which is not restricted to the near critical ($\mu\approx 1$) regime: 
the Landau--Ginzburg
energy restricted to $\M_N$ is precisely proportional to $\mu^2-1$. 
Our asymptotic formula for it (in the well-separated regime) is obtained 
by assuming approximate
equality with $V_N^\mu$, an assumption which can only hold for $\mu$ close to
$1$; but the formula (\ref{Udef}) itself is not restricted to the near-critical
regime. Of course, the validity of the whole moduli space approximation becomes
questionable for $|\mu-1|$ large, but this is a separate issue.

\begin{figure}[t]
\centerline{\epsfysize=3truein
\epsfbox{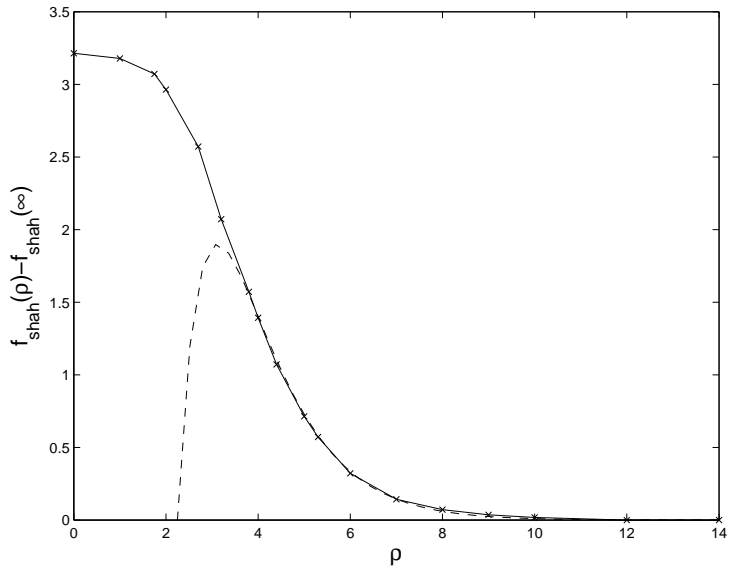}}
\caption{\it
The quantity $\int(1-|\phi|^2)^2d^2\xv$ for a vortex pair, as a function
of the vortex separation $\rho$. Solid line and crosses: Shah's numerical
data, taken from {\rm \cite{sha}}; dashed line: our conjectured asymptotic
formula, as in equation {\rm (\ref{fshah})}. 
}
\label{fig1}
\end{figure}

One way to test our conjecture is to
deduce from it an asymptotic formula for the integral
\beq
f_{\rm Shah}(\rho)=\int (1-|\phi|^2)^2d^2\xv
\eeq
for the case of critically coupled $N=2$ static vortices with intervortex
separation $\rho$ (note that $E(2)$ symmetry implies the integral may only
depend on $\rho$ in this case). The function $f_{\rm Shah}$ has been 
evaluated numerically by Shah \cite{sha}. Formula (\ref{Udef}) implies that
\beq
\label{fshah}
f_{\rm Shah}(\rho)-f_{\rm Shah}(\infty)=\frac{2q^2}{\pi}[\rho K_1(\rho)-
\nu K_0(\rho)]
\eeq
asymptotically at large $\rho$. A comparison of Shah's numerical data with
our conjectured asymptotic form is given in figure \ref{fig1}. The fit for
$\rho > 4$ is very good.

The $N$-vortex equations of motion in the asymptotic regime are obtained by
extracting the leading asymptotic term in the symplectic gradient of $H$.
Since $H$ is already exponentially small, an exponential correction to 
$\omega_{L^2}$ makes no contribution at this order, so we can actually set
$\omega_{L^2}=\pi\omega_0$, where 
\beq
\omega_0
=\frac{i}{2}\sum_r dz_r\wedge d\ol{z}_r
\eeq
is the canonical symplectic form
on $\C^N$. We find, to leading order,
\beq
\label{aeom}
\dot{z}_r=i
\sum_{s\neq r}F(|z_{rs}|)
z_{rs},\quad
F(\rho):=-\frac{\U'(\rho)}{2 \pi\gamma\rho}.
\eeq
Note that system (\ref{aeom}) is 
similar to the equations of motion for a system of identical fluid point
vortices, or geostrophic vortices, of vorticity $1/\gamma$, 
which would result from replacing $\U$ by
\beq
\label{alt}
\U_{\rm fluid}(\rho)=\log\rho,\quad
\U_{\rm geo}(\rho)=K_0(\rho)
\eeq
respectively \cite{flu,geo}. The system with $\U_{\rm fluid}$ is particularly
well studied. We shall see that our system behaves rather more like
the geostrophic vortex system, which has not been so heavily
studied, though there are still significant differences.

Two-vortex dynamics in the moduli space approximation is almost trivial: the
vortices orbit their centre of mass at constant speed and separation. This is
a special case of the rotating polygon solutions whose stability properties
we will analyze in section \ref{rotpol}. We turn now to a discussion of
3-vortex dynamics.

\section{The dynamics of three vortices}
\label{dynthrvor}
\news

Novikov has given a thorough treatment of the internal dynamics of 
identical vortex triples interacting via $\U_{\rm fluid}$ in \cite{nov}.
His method can be adapted readily to deal with our system in its asymptotic 
form, (\ref{aeom}). The basic idea is to identify trajectories in the 
internal phase space of the system with level curves of the quantities 
conserved by flow (\ref{aeom}).
Since Novikov's method relies on exploiting the conservation laws
enjoyed by the system, it is more satisfactory to apply it directly to the
momenta conserved by the full flow, in asymptotic form, rather than the
momenta conserved by the asymptotic flow
(\ref{aeom}), and this is how we shall proceed. Of course, either approach
yields the same results.

To begin with, let $N$ be general.
Let us define the {\em centroid} of an $N$-vortex configuration by 
\beq
Z:=\frac{1}{i\pi N}(P_1+iP_2)=
\frac{1}{N}\sum_r z_r 
\eeq
and note that
\beq
\sum_r|z_r|^2=N|Z|^{2}+ \frac{1}{2N}\sum_{r}\sum_{s\neq r}|z_{rs}|^2.
\eeq
Since $Z$ and $P_0$ are conserved, it follows that
\bea
Q & := & \frac{2N}{\pi}P_0 - N^2 |Z|^2  \nonumber \\
& = & \sum_{r} \left( 
N(b_r \ol{z}_r+\ol{b}_{r} z_r)-\frac{1}{2}\sum_{s\neq r}|z_{rs}|^2
\right)
\eea
is conserved also.
The corresponding hamiltonian vector field can be readily
computed as
\bea
\X_{Q}& = & \frac{2N}{\pi}X_{0}-\frac{1}{\pi^2}(2 P_1 X_1 + 2 P_2 X_2)
\nonumber \\
& = & \frac{2Ni}{\pi}\sum_r \left(
(z_r - Z)\frac{\cd}{\cd z_r}-(\ol{z}_r - \ol{Z})\frac{\cd}{\cd \ol{z}_r}
\right),
\eea
and this describes rigid rotations of vortex configurations about their
centroids. Using (\ref{id1}), (\ref{id2}) and (\ref{romid}), one finds
\beq \label{PjQcom}
\{ P_j, Q \}=0, \qquad j=1, 2. 
\eeq

For a moment, let us consider $N=3$. Equation (\ref{PjQcom}) implies 
that the set of three
conserved quantities $\{P_1, Q, H\}$ is in involution. If the corresponding
hamiltonian vector fields are linearly independent at all points of an open
dense subset $\M_3 \setminus S \subset \M_3$, 
the dynamics of three vortices is Liouville integrable~\cite{aud} on the
six-dimensional phase space $\M_3$. This manifold is then foliated by 
invariant tori~\cite{arn}, some of them possibly degenerate. 
Clearly, $X_1$ is independent of both $\X_{Q}$ and $\X_{H}$ everywhere 
in $\M_N$. Hence, the question of Liouville integrability reduces to the 
condition of linear independence of $\X_Q$ and $\X_H$. Our intuition about 
the dynamics of $N=2$ vortices~\cite{man} leads us to claim that this
condition is generic on $M_3$. Actually, we expect $\X_Q$ and $\X_H$ to 
be linearly dependent exactly at the points of a subset $S \subset \M_3$
consisting of configurations with special symmetry: $z_r$ either at the 
vertices of an equilateral triangle, at the ends and midpoint of a line 
segment, or on the locus of coincidence $\Delta_3$; 
notice that all these 
conditions are algebraic in the coordinates $z_r$ (and their complex 
conjugates), so that $S$ is indeed closed with dense complement.
We shall see in the following that this picture is consistent with what we 
can learn about the dynamics of 3-vortices in the asymptotic regime of large 
separation. The configurations in $S$ that correspond to well-separated 
3-vortices are special cases of the vortex polygons that we shall discuss in 
section~\ref{rotpol}.

When all $|z_{rs}|$ are large, the angular momentum becomes
\beq
P_0=\frac{\pi}{2}\left[\sum_r|z_r|^2+\frac{q^2}{2\pi^2}
\sum_{r}\sum_{s\neq r}
|z_{rs}|K_1(|z_{rs}|)\right],
\eeq
whereas
\beq
Q=\frac{1}{2}
\sum_r \sum_{s\neq r}\left[|z_{rs}|^2+\frac{Nq^2}{\pi^2}|z_{rs}|
K_1(|z_{rs}|)\right].
\eeq
Note that, unlike $Z$ and $P_0$, $Q$ and $H$ depend only on the relative 
separations of the vortices $|z_{rs}|$, to this order.

If we now set $N=3$ and
define $\xv=(|z_{23}|,|z_{31}|,|z_{12}|)$, we see that
$\xv(t)$ is confined to a level curve of $(\hat{H},\hat{Q})$ in $\R^3$, where
\bea
\hat{H}&=&\U(x_1)+\U(x_2)+\U(x_3),\nonumber \\
\hat{Q}&=&g(x_1)+g(x_2)+g(x_3),\quad g(x):=x^2+\frac{Nq^2}{\pi^2}xK_1(x).
\eea
Our strategy is, then, to construct such level curves.
This almost completely determines the internal dynamics of the triple
(meaning the dynamics up to rigid rotations), 
up to
reparametrization of time. The only thing not determined is a discrete
variable, the 
{\em orientation} of the triple. Two orientations are possible: traversing
the triangle with vertices at $z_r$ clockwise can order the vertices $123$ or
$132$.

To construct a level curve of $(\hat{H},\hat{Q})$, we may choose some 
initial data $\xv(0)=\xv_0$ and then solve the first order system
\beq
\label{level}
\dot{\xv}(s)=\frac{\nabla\hat{H}\times\nabla\hat{Q}}{
|\nabla\hat{H}\times\nabla\hat{Q}|}
\eeq
to yield a level curve in arc-length parametrization ($|\dot{\xv}(s)|\equiv
1$). Since we work only to linear order in exponentially small quantities, and
$\hat{H}$ is small, we may keep only the leading term in $\hat{Q}$, that is,
take
\beq
\hat{Q}=|\xv|^2
\eeq
in (\ref{level}). So level sets of $\hat{Q}$ may be approximated by concentric
spheres of radius $\hat{Q}^\frac{1}{2}$ in $\R^3$, and the level curves of
$(\hat{H},\hat{Q})$ are simply the intersection curves between level sets of
$\hat{H}$ and these spheres. Alternatively, given a radius 
$\hat{Q}^\frac{1}{2}$, we may seek level curves of $H^{\hat{Q}}:S^2\ra\R$,
\beq
H^{\hat{Q}}(\yv)=\U(\hat{Q}^\frac{1}{2}y_1)+
\U(\hat{Q}^\frac{1}{2}y_2)+
\U(\hat{Q}^\frac{1}{2}y_3),
\eeq
where $\xv=\hat{Q}^\frac{1}{2}\yv$ and $S^2$ is the unit 2-sphere in $\R^3$.
It is straightforward to generate contour plots of $H^{\hat{Q}}$ for given
$\hat{Q}$ within any convenient coordinate system on $S^2$, and hence deduce
the corresponding internal trajectories. Note that $\hat{Q}$ determines the
root-mean-square of the triangle side lengths, $x_1,x_2,x_3$. It turns out
that the level curves are not very sensitive to $\hat{Q}$, provided we choose 
it to be fairly large (which it should be, in the asymptotic regime), so let
us imagine that $\hat{Q}$ has been fixed.

\begin{figure}[t]
\centerline{\epsfysize=3truein
\epsfbox{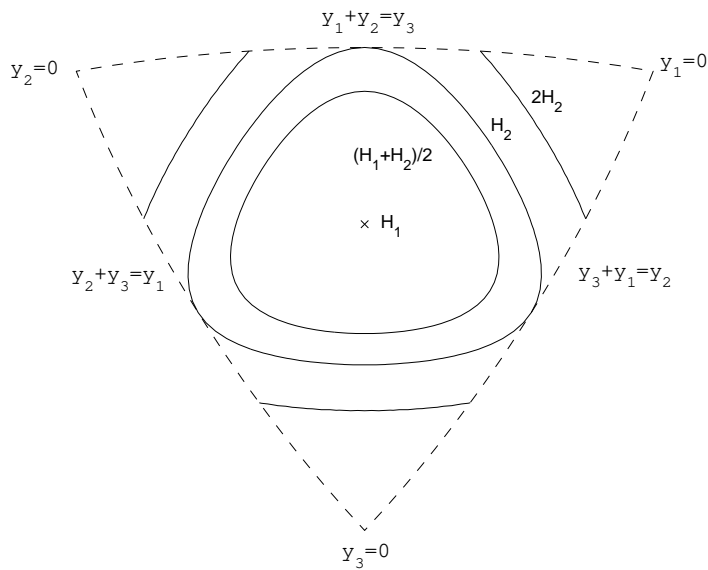}}
\caption{\it
Level curves of $\hat{H}^{\hat{Q}}$ for $\hat{Q}=441$. Here $H_1
=\hat{H}^{\hat{Q}}_{\rm min}$ and $H_2=\hat{H}^{\hat{Q}}_{\rm crit}$. The
dashed curves represent the colinearity boundaries, which intersect at
coincidence points. 
}
\label{fig2}
\end{figure}

Not every $\yv\in S^2$ represents a valid vortex triple. Each $y_i$ must be
non-negative, and since they represent the side lengths of 
a euclidean triangle, none
may exceed the sum of the other two. So $\yv(t)$ is confined to a 
triangular region in $S^2$, bounded by great-circular arcs
(the intersection of the sphere with
the colinearity planes $y_1=y_2+y_3$, $y_2=y_3+y_1$, $y_3=y_1+y_2$), 
and centred
on $\yv_0=(1,1,1)/3^\frac{1}{2}$. This central point
represents an equilateral triple.
The orthogonal projection of this region onto the tangent plane $T_{\yv_0}S^2$
is depicted in figure \ref{fig2}. The boundaries project to congruent
elliptic arcs. Any point on the boundary represents a colinear vortex triple,
and the boundaries intersect at coincidence points (where a vortex pair 
coalesces). Since our analysis is restricted to the well-separated regime, 
level curves which approach these corners closely are of suspect validity.

Figure \ref{fig2} depicts level curves for the choice $\hat{Q}=441$ and
several values of $\hat{H}$. Our discussion is qualitatively similar to
Novikov's \cite{nov}, so will be kept brief. 
The minimum possible value of $H^{\hat{Q}}$ is
\beq
H_{\rm min}^{\hat{Q}}=H^{\hat{Q}}(\yv_0)=3 
\, \U(\sqrt{\hat{Q}/3})
\eeq
attained by the equilateral triple. The level ``curve'' for this value consists
of the single point $\yv_0$, so the internal configuration of this triple
remains constant. In fact the triangle rotates about its centre at constant
speed, as will be discussed in section \ref{rotpol}. Increasing $H^{\hat{Q}}$,
we obtain closed curves centred on $\yv_0$. In this low energy regime, the
shape of the triangle varies periodically in time as the triangle spins, but
the triangle's orientation remains constant. The periods of shape change and
rotation are generically incommensurate. Clearly, these solutions are all 
stable.

Increasing $H^{\hat{Q}}$ further, the level curve eventually becomes tangent to
the colinearity boundaries, at the critical value
\beq
H_{\rm crit}^{\hat{Q}}=H^{\hat{Q}}((1,1,2)/6^\frac{1}{2})= 2\, 
\U(\sqrt{\hat{Q}/6}) +\U(2\sqrt{\hat{Q}/6}).
\eeq
This level curve should not be interpreted as a closed orbit.  Rather,
we have
three
 fixed points, $(1,1,2)/\sqrt{6}$, $(1,2,1)/\sqrt{6}$, $(2,1,1)/\sqrt{6}$, 
joined by 3 heteroclinic orbits. The fixed points represent colinear triples
with the middle vortex equidistant from the outer vortices, a special
case of the filled rotating polygons which will be discussed in section
\ref{filrotpol}. This is stationary (but not static) and clearly unstable: 
small perturbations lead to divergent dynamical regimes. 

For $H^{\hat{Q}}>H_{\rm crit}^{\hat{Q}}$,
 the level curve splits into 3 disjoint
pieces. The actual trajectory stays on one segment of this curve -- it reverses
direction each time it hits a colinearity boundary, with a corresponding 
change in the triangle's orientation. At these points, the triangle collapses
and turns itself inside out, so for larger energies, the triple's internal 
motion consists of a periodic ``flip-flopping''. As the energy grows unbounded,
the trajectory approaches a coincidence point, and hence escapes the 
presumed region of validity of our approximation. 

It is interesting to note that {\em every} trajectory intersects the isosceles
lines (the straight lines containing $\yv_0$ and the coincidence points),
so every 3-vortex motion passes, at least once, through an isosceles 
configuration. 

\section{Rigidly rotating polygons}
\label{rotpol}
\label{filrotpol}

\news

If $N$ vortices are placed at the vertices of a regular $N$-gon, they will
rotate at constant speed about their 
centroid, which remains fixed \cite{man}. 
This behaviour is familiar from both the fluid and geostrophic vortex systems. 
Given the asymptotic equations of motion (\ref{aeom}), it is straightforward 
to derive frequency-radius relations and spectral stability properties of
these solutions, valid for large polygon side length.
Note that a change in the coupling constants $\mu$, $\gamma$ may always be 
absorbed into a rescaling and/or reversal of time, so the stability properties
will be independent of our choice of parameters.

It is convenient to transform
to a co-rotating frame by defining new coordinates $y_r(t)\in\C$,
\beq
z_r(t) =: y_r(t)e^{i\Omega t}
\eeq
for some fixed choice of frequency $\Omega\in\R$, so that frequency $\Omega$
solutions are static in the new frame. System (\ref{aeom}) in these 
coordinates is
\beq
\label{ceom}
\dot{y}_r=i\sum_{s\neq r}F(|y_{rs}|)y_{rs}-i\Omega y_r,
\eeq
where $y_{rs}:=y_r-y_s$. Note that (\ref{ceom}) is hamiltonian flow on
$(\C^N,\omega_0)$ with respect to the modified Hamiltonian
\beq
H_{\Omega}=
\sum_r \sum_{s\neq r}\U(|y_{rs}|)-\Omega\sum_r|y_r|^2.
\eeq
Let $\lambda :=e^{2\pi i/N}$. 
Then $y_r=\lambda^r\sigma$, $\sigma\in(0,\infty)$,
is a static solution of (\ref{ceom}) provided
\beq
\label{radom}
\Omega=\sum_{j=1}^{N-1}F(|1-\lambda^j|\sigma).
\eeq
This is the frequency-radius relation for a rotating $N$-gon.

We turn now to stability properties. Let $y\in\C^N$ be a fixed point of the
co-rotating flow. Then we may define the linearization of the flow about $y$,
\beq
\Lambda_y:T_y\C^N\ra T_y\C^N,\qquad \Lambda_y:X\mapsto\nabla_X \X_{H_\Omega},
\eeq
where $\nabla$ is 
{\em any} affine connexion on $\C^N$, and $\X_{H_\Omega}$ is the 
symplectic gradient of $H_\Omega$. The fixed point $y$ is spectrally stable
if $\Lambda_y$ has no eigenvalues with positive real part. Now, since 
$\X_{H_\Omega}$ is symplectic, it follows that 
$\Lambda_y\in\aut(T_y\C^N,\omega_0)$, that is,
\beq
\omega_0(\Lambda_y X,Y)+\omega_0(X,\Lambda_y Y)\equiv 0
\eeq
for all $X,Y \in T_y\C^N$. 
It follows that, if $\eta$ is an eigenvalue of $\Lambda_y$, so are $-\eta$,
$\ol{\eta}$ and $-\ol{\eta}$. In order that $y$ be spectrally stable, 
therefore, $\spec\Lambda_y$ must be purely imaginary.

In the case of the rotating $N$-gon, $y=\tilde{y}=
(\lambda\sigma,\lambda^2\sigma,\ldots,\sigma)$, one can find 
$\spec\Lambda_{\tilde{y}}$ explicitly due to the cyclic symmetry of the
configuration \cite{geo}. Let $\pi:\Z\ra\Z_N$ be the group homomorphism with
$\pi(1)=1$. Then a $N\times N$ real matrix $A$ is {\em right circulant} if
there exists $\hat{A}\in\R^N$ such that
\beq
A_{ij}=\hat{A}_{\pi(j-i+1)}
\eeq
for all $i,j$. In order to exploit this symmetry, it is convenient to define
a twisted complex basis $e_{(i)}\in\C^N$, $i=1,\ldots,N$, such that
\beq
\label{twistdef}
e_{(i)j}=\lambda^i\delta_{ij}
\eeq
there being no summation implied. If we write down the matrix representing
$\Lambda_{\tilde{y}}$ with respect to the associated basis for $\R^N$, that is,
$\{\re e_{(1)},\ldots,\re e_{(n)},\im e_{(1)},\ldots,\im e_{(n)}\}$, we find
that
\beq
\Lambda_{\tilde{y}}=\left[\begin{array}{cc}A&B\\C&A\end{array}\right]
\eeq
where each of the real $N\times N$ blocks is right circulant. The explicit 
formulae for $\hat{A},\hat{B},\hat{C}\in\R^N$ are rather complicated, and are
presented in the appendix. Of more interest are the following symmetries
possessed by these coefficients:
\beq
\label{refsym}
\left.\begin{array}{rcr}\hat{A}_1=0,\quad \hat{A}_{N+2-s}&=&-\hat{A}_s\\
\hat{B}_{N+2-s}&=&\hat{B}_s\\
\hat{C}_{N+2-s}&=&\hat{C}_s\end{array}\right\}
s=2,\ldots,N.
\eeq

Right circulancy is a powerful symmetry because the eigenvectors of a right 
circulant matrix are independent of its entries. For each 
$j\in\{1,\ldots,N\}$,
let
\beq
\label{xdef}
\xv_j=(1,\lambda^j,\lambda^{2j},\ldots,\lambda^{(N-1)j})\in\C^N.
\eeq
Then $\xv_j$ is simultaneously an eigenvector of $A$, $B$ and $C$, with
eigenvalues
\beq
\label{spec1}
\alpha_j=\hat{A}\cdot\xv_j,\quad
\beta_j=\hat{B}\cdot\xv_j,\quad
\gamma_j=\hat{C}\cdot\xv_j.
\eeq
From (\ref{refsym}), (\ref{xdef}) and (\ref{spec1}), one sees that 
$\alpha_j$ is purely imaginary, while $\beta_j$ and $\gamma_j$ are real.
Note that $\ol{\xv}_j=\xv_{N-j}$ for all $1\leq j\leq N-1$, and that
$\ol{\xv}_N=\xv_N$. It follows that
\beq
\label{specsym}
\alpha_{N-j}=\ol{\alpha}_j=-\alpha_j,\quad
\beta_{N-j}=\ol{\beta}_j=\beta_j,\quad
\gamma_{N-j}=\ol{\gamma}_j=\gamma_j,\quad
\eeq
for $1\leq j\leq N-1$, and that $\alpha_N=\ol{\alpha}_N=-\alpha_N$, so
$\alpha_N=0$.
Short calculations using the detailed forms of $\hat{B}$, $\hat{C}$
(see appendix) establish that
$\beta_N=0$ and $\gamma_1=-\beta_1$.

Now, relative to the basis $\{\xv_1^+,\xv_1^-,\ldots,\xv_N^+,\xv_N^-\}$,
where $\xv_j^\pm:=(\xv_j,\pm\xv_j)\in\C^{2N}$, $\Lambda_{\tilde{y}}$ is 
block diagonal, with $2\times 2$ blocks
\beq
\Lambda_j=\left[\begin{array}{cc}\alpha_j+\half(\beta_j+\gamma_j)&
\half(\beta_j+\gamma_j)\\
-\half(\beta_j-\gamma_j) &
\alpha_j-\half(\beta_j+\gamma_j)\end{array}\right].
\eeq
It follows that the eigenvalues of $\Lambda_{\tilde{y}}$ are
\beq
\eta_j^\pm=\alpha_j\pm\sqrt{\beta_j\gamma_j}.
\eeq
Owing to the symmetry properties of $\alpha_j$, $\beta_j$, $\gamma_j$,
(\ref{specsym}), we see that 
\beq
\eta_{N-j}^\pm\equiv-\eta_j^\pm,
\eeq
so that eigenvalues generically come in quartets. Note also that
$\eta_N^\pm=0$ (since $\alpha_N=\beta_N=0$) and that $\eta_1^+=-\eta_{N-1}^-=
i\Omega$. These eigenvalues originate
from the $E(2)$ symmetry enjoyed by system (\ref{ceom}), as will be shown
shortly.

Since $\alpha_j\in i\R$ for all $j$, $\tilde{y}$ is spectrally stable provided
that $\beta_j\gamma_j\leq0$ for all $j$. But $\beta_1\equiv -\gamma_1$, and
$\beta_{N-j}\gamma_{N-j}\equiv\beta_j\gamma_j$ for $2\leq j\leq N-1$, so
there are in fact only $k-1$ stability criteria, where $N=2k$ or $N=2k+1$, 
namely
\beq
\beta_j\gamma_j\leq 0,\qquad j=2,\ldots,k.
\eeq
For $N=2,3$ stability is automatic, for $N=4,5$ there is one criterion,
for $N=6,7$ two criteria, and so on.

In the first nontrivial case, $N=4$, 
$\lambda=i$ and the single stability
criterion is
\beq
(\beta_2\gamma_2)(\sigma)=-8\sqrt{2}\sigma^2 F'(\sqrt{2}\sigma)F'(2\sigma)
\leq 0.
\eeq
Since $F$ has no critical points in the well-separated (large $\sigma$)
regime, the rotating square is spectrally stable for all $\sigma$
sufficiently large. As $N$ increases, the expressions for $\beta_j\gamma_j$
become increasingly complicated so that it is not feasible to check their
sign by hand. It is straightforward to check the criteria graphically by
plotting the sign of $\beta_j\gamma_j$ against $\sigma$, however. The results
are summarized and compared with the previously studied cases of fluid and
geostrophic vortices
in table \ref{tab1}.

\begin{table}
\begin{center}
\begin{tabular}{|c|ccccccc|}\hline
$N$            &2&3&4&5&5&7&$\geq 8$ \\ \hline
gauged         &1&1&1&1&0&0&0\\
fluid          &1&1&1&1&1&1&0\\
geostrophic    &1&1&1&1&0&0&0\\ \hline
\end{tabular}
\end{center}
\caption{\it Comparison of spectral stability properties for rotating vortex 
$N$-gons of various types, $0=$unstable,
$1=$stable. In the gauged and geostrophic cases, the entries refer to
stability at large polygon radius $\sigma$. There may be windows of anomalous
(in)stability at small $\sigma$, but these lie outside the regime of 
validity of the calculations. Entries for geostrophic and fluid vortices
are taken from {\rm \cite{geo}}.}
\label{tab1}
\end{table}

We may give a similar analysis of the case of $N-1$ vortices on the vertices
of a regular polygon orbiting another vortex at the polygon's centre. Let
$\lambda=e^{2\pi i/(N-1)}$. Then $\hat{y}=(0,\lambda\sigma,\lambda^2\sigma,
\ldots,\sigma)\in\C^N$ is a fixed point of the flow (\ref{ceom}), provided
that
\beq
\Omega=F(\sigma)+\sum_{j=1}^{N-2}F(|1-\lambda^j|\sigma)(1-\lambda^j).
\eeq
Once again the linearization $\Lambda_{\hat{y}}\in\aut(T_{\hat{y}}\C^N,
\omega_0)$, so $\hat{y}$ is spectrally stable if and only if
$\spec\Lambda_{\hat{y}}\subset i\R$. The presence of the centre vortex destroys
the right circulant symmetry of $\Lambda$, however, so we must resort to a
numerical algorithm to generate $\spec\Lambda_{\hat{y}}$.  By plotting
$\max\{|\re\eta|:\eta\in\spec\Lambda_{\hat{y}}\}$ against $\sigma$, we can
determine which of the $(N-1)$-gons are spectrally stable in the well
separated regime. The results, which were generated using Matlab's
eigenvalue finder, are summarized in table \ref{tab2}.

\begin{table}
\begin{center}
\begin{tabular}{|c|ccccccccc|}\hline
$N-1$          &2&3&4&5&5&7&8&9&$\geq 10$ \\ \hline
gauged         &0&1&1&1&1&1&0&0&0\\
fluid          &0&1&1&1&1&1&1&1&0\\
geostrophic    &0&0&1&1&1&1&0&0&0\\ \hline
\end{tabular}
\end{center}
\caption{\it Comparison of spectral stability properties for vortex 
$(N-1)$-gons rotating about a single central vortex for various vortex
types, key and comments as for table \ref{tab1}.}
\label{tab2}
\end{table}

These results have been derived in the approximation of large $\sigma$. In
the full adiabatic approximation, they will receive corrections
due to the subleading terms in $\omega_{L^2}$ and $H$, and 
we are assuming that these corrections
will be very small when $\sigma$ is large. What reason do we have to believe
that the seemingly delicate property of spectral stability,
$\spec\Lambda\subset i\R$, will not be destroyed by these perturbations?
The spectra found exactly (for $\Lambda_{\tilde{y}}$), or numerically
(for $\Lambda_{\hat{y}}$), had no accidental degeneracies for $\sigma$ 
sufficiently large, in the stable cases. All the eigenvalues were
simple, with the exception of $\eta=0$, which we will shortly return to.
No simple imaginary eigenvalue can be perturbed off the imaginary axis because,
due to the reflexion symmetries enjoyed by $\spec\Lambda$, it would 
have to split in two, which is impossible by conservation of multiplicity.
Further, the $0$ eigenvalue, which turns out to have multiplicity $2$, 
is fixed at $0$ by symmetry considerations, as we  shall now see.

We are free to transform to the co-rotating frame in the full
adiabatic flow, by redefining our Hamiltonian
\beq
H\mapsto H_\Omega=H-\Omega P_0.
\eeq
Then $\tilde{y}$ and $\hat{y}$ will still be fixed points of $\X_{H_\Omega}$,
provided $\sigma$ is chosen suitably. In each case ($y=\tilde{y}$ or $\hat{y}$)
we may study the
spectrum of the linearization $\Lambda_y:X\mapsto\nabla_X\X_{H_\Omega}\in
\aut(T_y\M_N,\omega_{L^2})$, which will be a slightly perturbed version of the
asymptotic spectrum already considered. 
Recall that $X_0$, $X_1$, $X_2$  denote the vector fields generating rotations
and translations (\ref{Xdef0})--(\ref{Xdef2}).
Now for any fixed point $y\in\M_N$,
\beq
\Lambda_yX_j=\nabla_{X_j}\X_{H_\Omega}=[X_j,\X_{H_\Omega}]=-\Omega[X_j,X_0]
\eeq
since $[X_j,\X_H]=\X_{\{P_j,H\}}=0$ by the $E(2)$ symmetry. Hence
\beq
\Lambda_y X_0=0,\quad\Lambda_y X_1=-\Omega X_2,\quad\Lambda_y X_2=\Omega X_1,
\eeq
whence it follows that $\{0,i\Omega,-i\Omega\}\subset\spec\Lambda_y$. Hence,
the double $0$ eigenvalue is fixed under any perturbation maintaining $E(2)$
symmetry, so we conclude that the simple stability analysis given above is
structurally stable at large $\sigma$. 

\section{Conclusion}
\news
\label{con}

In this paper we have confirmed the expectation raised by 
\cite{rom} that the adiabatic approximation to $N$-vortex dynamics in Manton's
model of a planar superconductor amounts to a natural hamiltonian flow on
$(\M_N,\omega_{L^2})$. A hamiltonian account of the conserved
momenta descending from the symplectic but nonhamiltonian action of $E(2)$ on
$\M_N$ has been given, 
and based on it we argued that the dynamics of three
vortices is integrable in the sense of Liouville. 
We have derived an asymptotic formula for the 2-vortex interaction
potential close to critical coupling, namely
\beq
\U(\rho)=\frac{q^2}{4\pi}(\mu^2-1)[\rho K_1(\rho)-\nu K_0(\rho)],\qquad
q=-2\pi 8^\frac{1}{4},\quad \nu\approx 2.7060,
\eeq
and used this to analyze internal 3-vortex dynamics at large separation.
We have studied the spectral stability properties of rigidly rotating vortex
polygons, both with and without a central vortex, and compared our results
with the more familiar cases of fluid and geostrophic vortices.

Now that we have some quantitative dynamical predictions from the adiabatic
approximation, it would be interesting to test its validity numerically. There
are really two separate issues. First, does the hamiltonian flow give
a good account of low-energy vortex dynamics near critical coupling? Second,
do our asymptotic formulae for $\U$ and $\omega_{L^2}$ give a good 
approximation to this flow at large separation? Both questions may be addressed
by numerical analysis of 2-vortex dynamics. According to the adiabatic
approximation, two vortices orbit one another at constant radius and angular
frequency indefinitely. Is this behaviour seen in numerical simulations of the
full field theory? One would expect the orbiting pair to gently radiate energy
and hence slowly drift either together or apart (depending on the choice of 
$\mu$, $\gamma$), but hopefully this happens on a much longer time scale than
the period of their orbit. If their separation $\rho$ is large, then their
angular frequency should be close to
\beq
\Omega(\rho)=F(\rho)=\frac{q^2}{8\pi\gamma}(\mu^2-1)\left(1+\frac{\nu}{\rho}
\right)K_0(\rho).
\eeq
This is very different from the inverse square law found for ungauged 
Landau--Ginzburg vortices \cite{lansch}. It would be interesting to test this
formula numerically in terms of both its $\rho$ and $\mu$ dependence.

\section*{Acknowledgements}

The authors wish to thank Nick Manton for useful discussions and
correspondence.

\section*{Appendix}
\renewcommand{\theequation}{A\arabic{equation}}
\news

The simplest way to construct the matrix representing $\Lambda_{\tilde{y}}$ 
with respect to the twisted basis $e_{(i)}$ defined in equation 
(\ref{twistdef})
is to incorporate the twisting into the 
co-rotating coordinate system, by
defining $x_r(t)$ such that $y_r(t)=\lambda^r x_r(t)$. The equations of
motion (\ref{ceom}) become
\beq
\label{a1}
\dot{x}_r=i\sum_{s\neq r}F(|x_r-\lambda^{s-r}x_s|)(x_r-\lambda^{s-r}x_s)-
i\Omega x_r.
\eeq
Then $\tilde{x}=(\sigma,\sigma,\ldots,\sigma)$ is a fixed point of 
(\ref{a1}) provided (\ref{radom}) holds. The linearization of (\ref{a1}) about
$\tilde{x}$ is
\beq
\label{a2}
\delta\dot{x}_r=iR(\sigma)\delta x_t-i S(\sigma)\delta\ol{x}_r
-i\sum_{s\neq r}\lambda^{s-r}Q(|1-\lambda^{s-r}|\sigma)\delta x_s
+i\sum_{s\neq r}P(|1-\lambda^{s-r}|\sigma)\delta\ol{x}_s,
\eeq
where
\bea
P(\sigma)&:=&\half\sigma F'(\sigma), \nonumber \\
Q(\sigma)&:=&F(\sigma)+P(\sigma), \nonumber \\
R(\sigma)&:=&\sum_{j=1}^{N-1}
[\lambda^j F(|1-\lambda^j|\sigma)+P(|1-\lambda^j|\sigma)],\nonumber \\
\label{a3}
S(\sigma)&:=&\sum_{j=1}^{N-1}\lambda^j P(|1-\lambda^j|\sigma).
\eea
Note that $P,Q,R,S$ are all real. Decomposing (\ref{a2}) into real and 
imaginary parts, with $\delta x_r=\delta x_r^1+i\delta x_r^2$, one finds that
\beq
\label{a4}
\left[\begin{array}{c}
\delta\dot{x}_1^1\\ \vdots \\ \delta\dot{x}_N^1 \\ \delta\dot{x}_1^2 \\ \vdots
\\ \delta\dot{x}_N^2\end{array}\right]=
\left[\begin{array}{ccc}A& & B\\ & & \\ C& & A \end{array} \right]
\left[\begin{array}{c}
\delta{x}_1^1\\ \vdots \\ \delta{x}_N^1 \\ \delta{x}_1^2 \\ \vdots \\
\delta{x}_N^2\end{array}\right]
\eeq
where $A,B,C$ are the $N\times N$ right circulant matrices generated by
$\hat{A}$, $\hat{B}$, $\hat{C}\in\R^N$:
\beq
\label{a5}
\begin{array}{lll}
\hat{A}_1= 0 & \hat{A}_s = \im(\lambda^{s-1})Q(|1-\lambda^{s-1}|\sigma)&
s  = 2,\ldots,N  \\
\hat{B}_1 = -R(\sigma)-S(\sigma) & \hat{B}_s = 
\re(\lambda^{s-1})Q(|1-\lambda^{s-1}|\sigma)+P(|1-\lambda^{s-1}|\sigma) &
s = 2,\ldots,N   \\
\hat{C}_1  =  R(\sigma)-S(\sigma) & \hat{C}_s  = 
-\re(\lambda^{s-1})Q(|1-\lambda^{s-1}|\sigma)+P(|1-\lambda^{s-1}|\sigma)&
s  = 2,\ldots,N 
\end{array}
\eeq
The matrix in equation (\ref{a4}) is $\Lambda_{\tilde{y}}$ in this basis.
The symmetries of the components of $\hat{A},\hat{B},\hat{C}$ claimed in
(\ref{refsym}) follow immediately, once we note that 
$\lambda^N=1$ and $\lambda^{-1}=\ol{\lambda}$ so that $\lambda^{(N+2-s)-1}=
\ol{\lambda}^{s-1}$ when $2\leq s\leq N$. Note that
\bea
\beta_N&=&\sum_{s=1}^N\hat{B}_s=-R(\sigma)-S(\sigma)+
\sum_{s=2}^N\left[
\re(\lambda^{s-1})Q(|1-\lambda^{s-1}|\sigma)+P(|1-\lambda^{s-1}|\sigma)
\right] \nonumber \\
&=&\sum_{j=1}^{N-1}\left[(\re (\lambda^j)-\lambda^j)(F(|1-\lambda^j|\sigma)+
P(|1-\lambda^j|\sigma)\right]=0,
\eea
as previously claimed. Note also that
\bea
\beta_1+\gamma_1&=& \hat{B}_1+\hat{C}_1+\sum_{j=1}^{N-1}(\hat{B}_{j+1}+
\hat{C}_{j+1})\lambda^j \nonumber \\
&=&-2S(\sigma)+2\sum_{j=1}^{N-1}\lambda^j P(|1-\lambda^j|\sigma)=0
\eea
so that $\gamma_1=-\beta_1$, as was claimed.


\begin{thebibliography}{99}

\bibitem{flu} H. Aref,
``Point vortex motions with a center of symmetry''
{\sl Phys.\ Fluids} {\bf 25} (1982) 2183--2187.

\bibitem{arn} V.I. Arnold, 
{\sl Mathematical Methods of Classical Mechanics}, 2nd edition, 
New York, Springer-Verlag, 1989.

\bibitem{aud} M. Audin, 
{\sl Spinning Tops}, 
Cambridge, Cambridge University Press, 1996.

\bibitem{guiste} V. Guillemin and S. Sternberg, 
{\sl Symplectic Techniques in Physics}, Cambridge, Cambridge 
University Press, 1984.

\bibitem{hanetal} T.H. Hansson, S.B. Isakov, J.M. Leinaas and 
U. Lindstr\"om, 
``Classical phase space and statistical mechanics of identical particles''
{\sl Phys. Rev. E} {\bf 63} (2001) 026102, {\tt quant-ph/0003121}.

\bibitem{hashoryer} M. Hassa\"\i ne, P.A. Horv\'athy and J.-C. Yera,
``Non-relativistic Maxwell--Chern--Simons vortices''
{\sl Ann. Phys.} {\bf 263} (1998) 276--294, {\tt hep-th/9706188}.

\bibitem{jaftau} A. Jaffe and C. Taubes, 
{\sl Vortices and Monopoles}, Boston, Birkh\"auser, 1980.

\bibitem{lansch} O. Lange and B.J. Schroers,
``Unstable manifolds and Schr\"odinger dynamics of Ginzburg-Landau vortices''
{\sl Nonlinearity} {\bf 15} (2002) 1471--1488,
{\tt nlin.PS/0201047}.

\bibitem{man} N.S. Manton, ``First-order vortex dynamics''
{\sl Ann. Phys.} {\bf 256} (1997) 114--131, {\tt hep-th/9701027}.

\bibitem{mannas} N.S. Manton and S.M. Nasir,
``Conservation laws in a first-order dynamical system of vortices''
{\sl Nonlinearity} {\bf 12} (1998) 851--865, {\tt hep-th/9809071}.

\bibitem{manspe} N.S. Manton and J.M. Speight,
``Asymptotic interactions of critically coupled vortices''
{\sl Commun.\ Math.\ Phys.} {\bf 236} (2003) 535--555, {\tt hep-th/0205307}.

\bibitem{mcdsal} D. McDuff and D. Salamon, 
{\sl Introduction to Symplectic Topology}, 2nd edition, Oxford, 
Clarendon Press, 1998.

\bibitem{geo} G.K. Morikawa and E.V. Swenson,
``Interacting motion of rectilinear geostrophic vortices''
{\sl Phys.\ Fluids} {\bf 14} (1971) 1058--1073.

\bibitem{nov} E.A. Novikov,
``Dynamics and statistics of a system of vortices''
{\sl Sov. Phys. JETP} {\bf 41} (1975) 937--943.

\bibitem{rom} N.M. Rom\~ao, 
``Quantum Chern--Simons vortices on a sphere''
{\sl J. Math.\ Phys.} {\bf 42} (2001) 3445--3469, {\tt hep-th/0010277}.

\bibitem{romthe} N.M. Rom\~ao, 
{\sl Classical and Quantum Aspects of Topological Solitons}, 
PhD Thesis, University of Cambridge, 2002 (unpublished).

\bibitem{sam} T.M. Samols,  
``Vortex scattering''
{\sl Commun.\ Math.\ Phys.} {\bf 145} (1992) 149--179,

\bibitem{sha} P.A. Shah,
``Vortex scattering at near-critical coupling''
{\sl Nucl.\ Phys.} 
{\bf B294} (1994) 259--276, {\tt hep-th/9402075}.

\bibitem{spe} J.M. Speight,
``Static intervortex forces''
{\sl Phys.\ Rev.} {\bf D55} (1997) 3830--3835, {\tt hep-th/9603155}.

\bibitem{str} I.A.B. Strachan,
``Low-velocity scattering of vortices in a modified abelian Higgs model''
{\sl J. Math.\ Phys.} {\bf 33} (1992) 102--110.

\bibitem{stu1} D. Stuart, 
``Dynamics of abelian Higgs vortices in the near Bogomolny regime'' 
{\sl Commun.\ Math.\ Phys.} {\bf 159} (1994) 51--91.

\bibitem{ton} D. Tong, ``NS5-branes, T-duality and worldsheet instantons''
{\sl JHEP} {\bf 07} (2002) 
013--036, {\tt hep-th/0204186}.

\bibitem{tonhall} D. Tong, ``A quantum Hall fluid of vortices'', 
{\tt hep-th/0306266}.

\end{thebibliography}
\end{document}